\documentclass[leqno,twoside]{article}
\usepackage{amssymb,amsmath,latexsym, theorem}

\usepackage{amscd}

\title{A Note on Distributional Semi-Riemannian Geometry}
\author{Roland Steinbauer}
\date{November 2008}
\usepackage[dvipdfm, pdfstartview=FitH]{hyperref}
\usepackage{graphicx} 

\newtheorem{df}{Definition}[section]

\newtheorem{thr}{Theorem}[section]

\newcommand{\CC}{{\mathcal C}}
\newcommand{\Ll}{\ensuremath{L_{\mathrm{\scriptstyle loc}}}}
\newcommand{\Hl}{\ensuremath{H_{\mathrm{\scriptstyle loc}}}}
\newcommand{\Wl}{\ensuremath{W_{\mathrm{\scriptstyle loc}}}}
\newcommand{\R}{\mathbb R}\newcommand{\N}{\mathbb N}

\newcommand{\eps}{\varepsilon}
\newcommand{\al}{\alpha} 
\newcommand{\Om}{\Omega}
\newcommand{\om}{\omega}

\newcommand{\comp}{\subset\subset}
\newcommand{\supp}{\mathrm{supp}}

\newcommand{\D}{\ensuremath{{\mathcal D}}}

\newcommand{\G}{\ensuremath{{\mathcal G}}}

\newcommand{\cinfty}{{\cal C}^\infty}
\newcommand{\Cinfty}{\ensuremath{\mathcal C}^\infty}
\newcommand{\GT}{\Hl^1\cap\Ll^\infty}
\newcommand{\itemcr}{\hrule height 0pt width 40pt\hfill\break\vspace*{-\baselineskip}}

\newcommand{\g}{\ensuremath{\mathbf{g}}}
\newcommand{\X}{\mathfrak{X}}
\newcommand{\gs}{\ensuremath{{\mathcal G}}}

\newcommand{\esm}{\ensuremath{{\mathcal E}_M}}
\newcommand{\ns}{\ensuremath{{\mathcal N}} }

\newcommand{\Riem}{\mathrm{Riem}}
\newcommand{\Ric}{\mathrm{Ric}}
\newcommand{\GTconv}{\ensuremath{\Hl^1\cap\Ll^p}}

\begin{document}

\thispagestyle{empty}

\vspace{0.8cm}
\begin{center}
{\large \bf A note on distributional semi-Riemannian geometry\footnote{This work was partially 
supported by projects P20525 and Y237 of the Austrian Science Found (FWF).}
} \vspace*{3mm}

{\bf Roland Steinbauer\footnote{Department of Mathematics, University of Vienna, Nordbergstr.\ 15, A-1090
Wien, Austria, e-mail: roland.steinbauer@univie.ac.at}}
\end{center}

\begin{abstract}
We discuss some basic concepts of semi-Riemannian geometry
in low-regularity situations. In particular, we compare the settings of (linear) distributional 
geometry in the sense of L. Schwartz and nonlinear distributional geometry in the sense of
J.F. Co\-lom\-beau. 
\\[2mm] {\it AMS Mathematics  Subject Classification (2000)}: 
Primary: 83C75; 
secondary: 
46T30, 
53B30, 
46F10, 
46F30. 
\\[1mm] {\it Key words and phrases:} 
Semi-Riemannian metrics of low regularity, (nonlinear) distributional geometry, 
algebras of generalised functions.
\end{abstract}

\section{Introduction}

In this note we discuss some foundational concepts of semi-Riemannian
geo\-metry in case of low regularity. While usually semi-Riemannian geometry
is formulated for $\CC^\infty$-metrics, most of the results still hold true in case the
metric is locally $\CC^{1,1}$, i.e., its first derivatives being locally Lipschitz continuous: Indeed this
condition guarantees (local) unique solvability of the geodesic equation and implies
locally uniform boundedness of the curvature. In particular, the Riemann tensor can 
be interpreted as a distribution.

However, there is a strong motivation from physics to lower the regularity assumptions
on the metric. In particular, in the context of weakly singular space-times 
in general relativity such as thin shells of matter or radiation, cosmic strings, 
impulsive pp-waves, and shell crossing singularities one has to deal with Lorentz metrics 
of regularity below $\CC^{1,1}$. 

In this contribution we will mainly be concerned with the following two issues

\noindent 
(1) Defining the Levi-Civita connection of a metric of low regularity, and

\noindent 
(2) Defining the curvature from a connection or metric of low regularity,

\noindent 
in the context of two different mathematical frameworks, namely

\noindent (A) distributional geometry, i.e., the setting of tensor distributions 
in the sense of classical Schwartzian distribution theory, and

\noindent (B) nonlinear distributional geometry in the sense of Colombeau.

Approach (A) was pursued in \cite{marsden,gt,parker} and more recently in~\cite{LeFM}
building on global accounts to distribution theory e.g.\ provided in \cite{deRham}, while
approach (B) is due to~\cite{genprg} and based on global analysis (\cite{RD,ndg}) 
in (special) Colombeau algebras~(\cite{col}). 
Applications of (B) in general relativity can e.g.\ be found in~\cite{bal,clarke,scol,waveq}, 
and an overview of applications of (A) and (B) in relativity is provided by~\cite{SV}. 

While in this contribution we do not provide any new mathematics, we collect the 
respective results for both settings (A) and (B) and present them in parallel allowing 
for a direct comparison. Finally, we present results on the compatibility of (A) and (B) 
recently obtained in~\cite{gt-paper}.

In some more detail, the plan of this paper is as follows. 
After collecting the necessary prerequisites to make our presentation self-contained
in Sec.\ \ref{prerequisites}, in Sec.\ \ref{m+c} we define the 
notions of semi-Riemannian metrics and linear connections for each of the above frameworks (A) 
and (B). In Sec.\ \ref{fl} we deal with issue (1) and provide 
a version of the fundamental lemma of semi-Riemannian geometry for each of our settings,
while in Sec.\ \ref{curvature} we discuss issue (2) again for each of the frameworks (A) and (B). 
Finally, in Sec.\ \ref{compatibility} we answer the question of compatibility of the two 
approaches in the affirmative. Our main references for approach (A) and (B) will 
be~\cite{gt,LeFM} and \cite{genprg,gt-paper}, respectively.

\section{Linear and nonlinear distributional geometry}\label{prerequisites}
We recall that distributions on a smooth (paracompact, Hausdorff) mani\-fold $M$ 
of dimension $n$ are defined to be linear, continuous (w.r.t.\ the usual (LF)-topology) 
functionals on the space of compactly supported $n$-forms,
$\D'(M)=(\Om^n_c(M))'$. We will denote the action of a distribution on a test $n$-form
by $\langle v,\om\rangle$.
Distributional tensor fields and more generally distributional sections
of vector bundles can also be defined as elements of the dual space of appropriate 
spaces of sections. But for our purpose it will be sufficient (see however~\cite{mg}) 
to view them as tensor 
fields with distributional coefficients, or as $\CC^\infty(M)$-multilinear maps of 
vector fields and one-forms to scalar distributions, i.e., we have with $r,s$ denoting 
the tensor character
\begin{equation}\label{distgeo}
{\D'}^r_s(M)
=\D'(M) \otimes_{\cinfty(M)}{\mathcal T}^r_s(M)
\cong L_{{\mathcal C}^\infty(M)}\big(\Om^1(M)^r,\X(M)^s;\D'(M)\big).
\end{equation}
Here ${\mathcal T}^r_s(M)$ denotes the space of (smooth) $(r,s)$-tensor fields, and we 
have set ${\mathcal T}^1_0(M)=\X(M)$ and ${\mathcal T}^0_1(M)=\Om^1(M)$.
There is a well-developed theory of tensor distributions~(\cite{deRham,marsden,lich,parker}), 
which parallels the smooth case but suffers from the natural limitations of distribution theory.
In particular, in all multilinear operations only one factor may be distributional, 
while all others have to be smooth. For a pedagogical account we refer 
to~\cite[Ch.\ 3.1]{GKOS}.

One way to deal with products is to restrict oneself to subspaces of $\D'$. We will, 
in particular, be interested in Sobolev spaces. For $m\in\N_0$ and $1\leq p\leq\infty$ we
denote by $\Wl^{m,p}(M)$ the space of distributions  whose derivatives up to order $m$ 
locally belong to $L^p$.
Recall that $\Wl^{m,p}(M)$ is a Fr\'echet space with its topology induced by the 
semi-norms $\|u\circ\varphi_\alpha^{-1}\|_{W^{m,p}(V)}$, where $(U_\alpha,\varphi_\alpha)$
denotes the charts of an atlas for $M$, $V$ denotes any open, relatively compact 
subset of $\varphi_\alpha(U_\alpha)$, and $\|f\|_{W^{m,p}(V)}^p=\sum_{\al\leq m}\int_V
|\partial^\al f|^p $. Moreover, we write 
\[
(\Wl^{m,p})^r_s(M) = \Wl^{m,p}(M)\otimes_{\Cinfty(M)}{\mathcal T}^r_s(M)
\]
for the spaces of $\Wl^{m,p}$-tensor fields.
In case $p=2$ we use the usual convention and set $\Hl^m=\Wl^{m,2}$, 
and in case $m=0$ we obtain the usual local Lebesgue spaces which we denote by $\Ll^p$.

In nonlinear distributional geometry (\cite[Ch.\ 3]{GKOS}) in the sense
of J.F.\ Colombeau (\cite{col}) one replaces the vector space 
$\D'(M)$ by the algebra of generalised functions $\G(M)$ to
overcome the problem of multiplication of distributions.
Indeed, in the light of Schwartz' impossibility result (\cite{Schw1}),
this setting provides a minimal framework within which
tensor fields may be subjected to nonlinear operations, while maintaining 
consistency with smooth and distributional geometry: tensor products of 
smooth tensor fields are preserved as well as Lie derivatives of distributional ones.
The basic idea of the construction is smoothing of distributions (via
convolution) and the use of asymptotic estimates in terms of a
regularisation parameter: these are employed in a quotient construction which,
in particular, provides consistency with the product of smooth functions. 

The (special) Colombeau algebra of
generalised functions on $M$ is defined as the quotient
$$\gs(M) := \esm(M)/\ns(M)$$
of moderate nets of smooth functions modulo negligible ones, where
the respective notions are defined by ($P$ denoting linear differential operators
on $M$)
\[
\esm(M) :=\{ (u_\eps)_\eps\in\Cinfty(M):\, \forall K\comp M\,
\forall P\, \exists N:\, \sup\limits_{p\in K}|Pu_\eps(p)|=O(\eps^{-N}) \}\\
\]
\[\ns(M)  :=\{ (u_\eps)_\eps\in\Cinfty(M):\, \forall K\comp M\,
\forall P\, \forall m:\, \sup\limits_{p\in K}|Pu_\eps(p)|=O(\eps^{m}) \}.
\]
Elements of $\gs(M)$ are denoted by $u = [(u_\eps)_\eps] =
(u_\eps)_\eps + \ns(M)$. With componentwise operations, $\gs(\_)$ is a
fine sheaf of differential algebras where the derivations are Lie
derivatives with respect to smooth vector fields defined by
$L_Xu :=[(L_X u_\eps)_\eps]$, also denoted by $X(u)$. 

The $\gs(M)$-module ${\gs}^r_s(M)$ of generalised tensor fields can be defined
along the same lines using analogous asymptotic estimates. However, for our 
purpose it will suffice to set
\begin{eqnarray*}
\lefteqn{\gs^r_s(M)\,:=\,\gs(M)\otimes_{\Cinfty(M)}{\mathcal T}^r_s(M)}\\
&\cong&L_{\Cinfty(M)}(\Om^1(M)^r,\X(M)^s;\gs(M))
\cong L_{\gs(M)}(\gs^0_1(M)^r,\gs^1_0(M)^s;\gs(M)).
\end{eqnarray*}
Note that in contrast to classical distributions (c.f.\ (\ref{distgeo})),
generalised tensor fields map generalised (and not merely smooth) fields and forms 
to generalised functions. It is precisely this
property that allows one to raise and lower indices with the help of a
generalised metric (see Sec.\ \ref{m+c} below), just as in the smooth case.

Smooth functions are embedded into $\gs(M)$
simply by the \lq\lq constant\rq\rq\ embedding $\sigma$, i.e.,
$\sigma(f) := [(f)_\eps]$. In case $M\subseteq\R^n$ open, compactly
supported distributions are embedded into $\gs$ via convolution with
a mollifier $\rho \in {\mathcal S}(\R^n)$ with unit integral
satisfying $\int \rho(x) x^\alpha dx = 0$ for all $|\alpha| \geq 1$;
more precisely setting $\rho_\eps(x) = (1/\eps^n) \rho(x/\eps)$, we
define $\iota(w) := [(w * \rho_\eps)_\eps]$. In case $\supp(w)$ is 
not compact, one uses a sheaf-theoretical construction
which can be lifted to an arbitrary manifold using a partition of unity
subordinate to the charts of some atlas (\cite[Thm.\ 3.2.10]{GKOS}).
From the explicit formula, it is clear that the embedding commutes
with differentiation. It is, however, not
canonical since it depends on the mollifier as well as the partition
of unity. (A canonical embedding of distributions
\emph{is\/} provided by the so-called full version of the
construction (see~\cite{GlobTh,GlobTh2}), however, at the price of
a technical machinery, which we have chosen to avoid here.)

The interplay between generalised functions and distributions is most
conveniently formalised in terms of the notion of
association. We call a distribution $v \in \D'(M)$ associated with $u\in\gs(M)$ 
and write $u \approx v$ if, for all compactly supported $n$-forms $\om$ and one 
(hence any) representative $(u_\eps)_\eps$, we have
$\lim_{\eps \to 0} \int_M u_\eps \om = \langle w,\om\rangle$.

\section{Semi-Riemannian metrics and connections}\label{m+c}
Here we discuss Semi-Riemannian metrics and linear connections in the 
distributional and the generalised setting. To begin with we define
following Marsden (\cite[Def.\ 10.6]{marsden}).
\begin{df} A distributional $(0,2)$-tensor field $\g\in{\D'}^0_2(M)$
is called a distributional metric if it is symmetric and nondegenerate in the 
sense that $g(X,Y)=0$ for all $Y\in\X(M)$ implies $X=0\in\X(M)$.
\end{df}

Observe that due to its non-locality this condition of nondegeneracy is rather weak.
E.g.\ the classically singular line element $ds^2=x^2\,dx^2$ is nondegenerate in the 
above sense. Therefore it is appropriate to additionally ask for Parker's condition (\cite{parker})
demanding that $\g$ is nondegenerate in the usual sense off its singular support, see also
the discussion in \cite[Sec.\ 3]{gt-paper}.

By the above mentioned natural limitations of distribution theory it is not possible
to insert $\D'$-vector fields into $\g$, hence it does not induce a map ${\D'}^1_0\to{\D'}^0_1$
and cannot be used to pull indices of distributional tensor fields.
Moreover, the map induced by $\g:$ $\X(M)\ni X\mapsto X^\flat:=\g(X,.)\in{\D'}^0_1(M)$ is
injective but clearly not surjective, and in general there is no way to define the 
inverse metric. Also, notions like the index or geodesics of a distributional metric are not (easily) defined.

Let us now turn to the generalised setting. Following~\cite[Def.\ 3.4]{genprg} we define 
in this case (omitting some technicalities concerning the index).
\begin{df}\label{genmetric}
A symmetric section $\g\in\G^0_2(M)$ is called a generalised
semi-Riemannian metric if $\det\g$ is invertible in the generalised sense, i.e.,
for any representative
$\left( \det( \mathbf{g}_{\eps}) \right)_\eps$ of $\det\mathbf{g}$ we have
\[
\forall K\comp M\ \exists m \in \N:\
\inf_{p\in K} |\det( \mathbf{g}_{\eps})(p)| \geq \eps^m.
\]
\end{df}

This notion of nondegeneracy can be characterised pointwise (using
generalised points, see~\cite[Sec.\ 2]{genprg}) and the following 
characterisation of generalised metrics captures the
intuitive idea of a generalised metric as a net of classical metrics
approaching a singular limit: $\mathbf{g}$ is a generalised metric iff
on every relatively compact open subset $V\subseteq M$ there exists a
representative $( \mathbf{g}_{\eps})_\eps$ of $\mathbf{g}$ such that,
for fixed $\eps$, $\mathbf{g}_{\eps}$ is a classical metric and its
determinant, $\det\g$, is invertible in the generalised sense. The
latter condition basically means that the determinant is not too
singular.

A generalised metric induces a $\gs(M)$-linear isomorphism from
$\gs^1_0(M)$ to $\gs^0_1(M)$. The inverse of this isomorphism gives
a well-defined element of $\gs^2_0(M)$, the inverse metric, which we
denote by $\mathbf{g}^{-1}$, with representative
$\left( \mathbf{g}_{\eps}^{-1} \right)_\eps$ (\cite[Props.\ 3.6, 3.9]{genprg}).

Next we turn to connections. To fix notations we recall that classically a connection 
is a map $\nabla:\X(M)\times\X(M)\to\X(M)$ satisfying
($X,X',Y,Y'\in\X(M)$, $f\in\Cinfty(M)$)
\begin{eqnarray*}
(\nabla_1)&& \nabla_{fX+X'}Y=f\nabla_XY+\nabla_{X'}Y \\
(\nabla_2)&& \nabla_X(fY+Y')=f\nabla_X Y+X(f)Y+\nabla_X Y'.
\end{eqnarray*}
We now define.
\begin{df}\label{con} \itemcr 
\begin{itemize}
\item [(i)] A distributional connection (\cite[p.\ 358]{marsden}\footnote{Note, however, the typo in the very definition.},\cite[Def.\ 3.1]{LeFM}) 
is a map $\nabla:\ {\mathfrak X}(M)\times{\mathfrak X}(M)\to{\D'}^1_0(M)$
satisfying $(\nabla_1)$, $(\nabla_2)$ for all $X,X',Y,Y'\in\X$, $f\in\CC^\infty$.
\item [(ii)] A generalised connection (\cite[Def.\ 5.1]{genprg}) is a map $\nabla:\ \gs^1_0(M)\times\gs^1_0(M)\to\gs^1_0(M)$
 satisfying $(\nabla_1)$, $(\nabla_2)$ for all $X,X',Y,Y'\in\gs^1_0$, $f\in\gs$.
\end{itemize}
\end{df}

Both versions extend to the full smooth resp.\ generalised tensor algebra by using 
the Leibniz rule and defining $\nabla_Xu:=X(u)$ for scalars.
Also, in both cases the standard coordinate formulae hold.

\section{Versions of the fundamental lemma}\label{fl}

In this section we discuss the question in which sense a distributional
resp.\ generalised metric defines a Levi-Civita connection. 
Recall that classically the Levi-Civita connection $\nabla$ of a smooth metric $\g$ is 
given as the unique connection which is metric and torsion free, i.e., satisfies
\begin{eqnarray*}
 (\nabla_3)&& \nabla\g=0\ 
 \Big(\Longleftrightarrow X\big(\g(V,W)\big)=\g(\nabla_XV,W)+\g(V,\nabla_XW)\Big)\\
 (\nabla_4)&& T(X,Y):=\nabla_XY-\nabla_YX-[X,Y]=0,
\end{eqnarray*}
and is characterised by the Koszul formula
\begin{eqnarray*}
 2\g(\nabla_XY,Z)
&=&X\big(\g(Y,Z)\big)+Y\big(\g(Z,X)\big)-Z\big(\g(X,Y)\big)\\
&&-\g(X,[Y,Z])+\g(Y,[Z,X])+\g(Z,[X,Y])\,=:\,F(X,Y,Z).
\end{eqnarray*}
Observe that in the distributional framework $(\nabla_3)$ cannot be formulated:
a distributional connection can only act on smooth tensor fields but not on the distributional 
metric and likewise, in the terms on the r.h.s.\ of the condition equivalent to $(\nabla_3)$
the distributional metric cannot act on the distributional vector fields $\nabla_XV$ and 
$\nabla_XW$. 

One way to circumvent this obstacle is (following \cite[Sec.\ 4]{LeFM}) to primarily use the Koszul 
formula. Observe that its r.h.s.\ $F(X,Y,Z)$ is defined for an arbitrary distributional metric 
and $X,Y,Z\in\X(M)$, and the standard calculation shows that $Z\mapsto F(X,Y,Z)$ is $\Cinfty(M)$-linear. Hence
\[\nabla^\flat_XY:\ Z\mapsto \frac{1}{2}\,F(X,Y,Z)\]
defines a distributional one-form. But recall that we cannot use the metric to turn it into  
a distributional vector field, as is done in the smooth case.
On the other hand, it is readily shown that $\nabla^\flat$ satisfies the properties
($X,Y,Z\in\X(M)$)
\begin{eqnarray*}
(\nabla_3)^\flat&\qquad&\nabla^\flat_XY-\nabla^\flat_YX-[X,Y]^\flat=0\\
(\nabla_4)^\flat&\qquad&X\big(g(Y,Z)\big)-\nabla^\flat_XY(Z)-\nabla^\flat_XZ(Y)=0,
\end{eqnarray*}
which lead LeFloch and Mardare to define.
\begin{df}
The distributional Levi-Civita connection of a distributional metric $\g$ is 
defined as the mapping $\nabla^\flat:\ \X(M)\times\X(M)\to{\D'}^0_1(M)$ given by
\[\nabla^\flat_XY(Z):=\frac{1}{2}\,F(X,Y,Z).\] 
\end{df}
Note however, that $\nabla^\flat$ is {\em not} a distributional connection in the sense of
definition~\ref{con}(i): only if $\g$ possesses additional regularity 
we may set $\nabla_XY:=\g^{-1}(\nabla^\flat_XY,.)$, which implies
$(\nabla_3)$ and $(\nabla_4)$. This, of course, holds true if $\g$ is smooth 
but also if the conditions
\begin{equation}\label{motivate gt}
\nabla^\flat_XY\in(\Ll^2)^0_1(M)\ \mbox{and}\ \g^{-1}\in(\Ll^\infty)^0_2(M)
\end{equation}
hold: we then have that $\nabla_XY\in(\Ll^2)^1_0(M)\subseteq{\D'}^1_0(M)$.

Turning now to the generalised setting, we observe that we may follow the classical proof
of the fundamental lemma and use the properties of the inverse of the generalised metric
to obtain (cf.\ \cite[Thm.\ 5.2]{genprg}).
\begin{thr}
For any generalised metric $\g\in\gs^0_2(M)$ there exists a unique generalised
connection $\nabla$ that is metric and torsion free, i.e., satisfies $(\nabla_3)$ and 
$(\nabla_4)$ for all $X,Y,Z\in\gs^1_0(M)$. It is called the generalised Levi-Civita connection 
of $\g$ and is characterised by the Koszul formula.
\end{thr}

\section{Curvature}\label{curvature} 

Again we start by recalling the standard formula to fix our notation.
In the smooth setting the Riemann tensor is given by ($X,Y,Z\in\X(M)$)
\begin{equation}\label{Riem}
\Riem(X,Y)Z:=\nabla_X\nabla_YZ-\nabla_Y\nabla_XZ-\nabla_{[X,Y]}Z.
\end{equation}
Beginning with the distributional case, we immediately observe that the terms involving second
derivatives cannot be defined: $\nabla_X$ does not act on a general $\nabla_YZ\in{\D'}^1_0$. 
To answer the question for which restricted class of distributional connections we {\em can } define 
the curvature we consider (following~\cite[Sec.\ 3.2]{LeFM}) distributional connections which take values
in a subspace of the distributional vector fields, 
\[
 \nabla:\ \X(M)\times\X(M)\to{\mathcal A}(M)\subseteq{\D'}^1_0(M),
\]
where ${\mathcal A}(M)$ is to be chosen in such a way that $\nabla$ can be extended 
to it, i.e.,
\begin{eqnarray*}
 \nabla:\ \X(M)\times{\mathcal A}(M)\to{\D'}^1_0(M)\ \mbox{via}\ 
 \nabla_XY(\Theta):=X\big(Y(\Theta)\big)-\nabla_X\Theta(Y),
\end{eqnarray*}
where $X\in\X$, $Y\in{\mathcal A}$, and $\Theta\in\Om^1$.
Now the term $X(Y(\Theta))\in{\D'}^1_0$, and the obvious choice to make the action of 
$\nabla_X\Theta\in{\mathcal A}(M)$ on $Y\in{\mathcal A}(M)$ well-defined is to set ${\mathcal A}(M)
=(\Ll^2)^1_0(M)$. Indeed then $\nabla_X\Theta(Y)\in(\Ll^1)^1_0(M)$ can be 
interpreted as a distributional vector field and we may define.
\begin{df}\itemcr
\begin{itemize}
\item [(i)] A distributional connection $\nabla$ is called an $\Ll^2$-connection if
$\nabla_XY\in(\Ll^2)^1_0(M)$ for all $X,Y\in\X(M)$.
\item [(ii)] The distributional Riemann tensor $\Riem$ of an $\Ll^2$-connection is defined
by the usual formula \eqref{Riem}.
\end{itemize}
\end{df}
Note that for any $\Ll^2$-connection also the Ricci tensor and the scalar curvature can be
defined.

We now turn to the question of assigning a curvature to a distributional metric. Guided by the
above consideration we aim at an $\Ll^2$-Levi-Civita connection. By \eqref{motivate gt}
we see that a sufficient condition is $\g\in(\Hl^1\cap\Ll^\infty)^0_2(M)$ and $|\det\g|\geq C>0$
almost everywhere on compact sets. In fact, the latter condition together with the 
$\Ll^\infty$-bound on $\g$ implies local boundedness of $\g^{-1}$ by the cofactor formula. 
Hereby we have essentially rediscovered the key-notion of R. Geroch and J. Traschen's 
paper~\cite{gt} (however, see~\cite{LeFM} and~\cite{gt-paper} for the nondegeneracy condition) 
and may define.
\begin{df}~\label{gtmetric}
We call a distributional metric $\g\in(\GT)^0_2(M)$ gt-regular if it is a semi-Riemannian metric
(of fixed index) almost everywhere.
A gt-regular metric is called nondegenerate if its determinant is locally uniformly bounded away
from zero, i.e.,\begin{equation*}
 \forall K\comp M\ \exists C:\ |\det\g(x)|\geq C>0\
 \mbox{almost everywhere on}\ K.
\end{equation*}
\end{df}

Observe that $\GT(M)$ is an algebra and that the invertible elements are precisely those
which are locally uniformly bounded away from zero. 
In particular, the inverse of a nondegenerate gt-regular metric
is again gt-regular and nondegenerate in the sense that $\det(\g^{-1})$ is locally uniformly
bounded away from zero. Also note the similarity of this notion of nondegeneracy with
the nondegeneracy condition employed for generalised metrics in Definition~\ref{genmetric}.

Moreover, observe that by~\eqref{motivate gt} the distributional
Levi-Civita connection of a nondegenerate gt-regular metric really is a distributional connection
in the sense of Definition~\ref{con}(i). Finally, our above discussion indicates how to prove the folowing result 
which was first obtained in~\cite{gt} by analysing local coordinate expressions and rederived 
in~\cite{LeFM} in a coordinate invariant way.
\begin{thr}
For a nondegenerate gt-regular metric the Riemann and Ricci tensor and the scalar
curvature are defined as distributions.
\end{thr}

Summing up Definition~\ref{gtmetric} provides sufficient conditions on a distributional 
metric that allow to
perform the most basic operations of semi-Riemannian geometry. On the other hand the 
question of necessity is hard to tackle in a precise sense, but there are strong indications
that we have indeed found the most general ``reasonable'' distributional framework providing the
geometric foundations of general relativity. First of all the Bianchi identities, which provide
conservation of energy, cannot be formulated in the gt-setting. Moreover, the
following consideration is essential when modelling singular scenarios 
in relativity: since a distributional metric does not directly make sense as a physical model
we have to interpret it as an idealisation obtained as the limit of some approximating sequence
of ``physically realistic'' metrics. It is now vital to have at hand a notion of convergence for 
these sequences that also implies convergence of the respective curvature quantities. 
While such stability properties have been derived for gt-regular metrics (see also Section~\ref{compatibility} below) it is known that such results are not available for a 
slightly wider class of metrics considered in~\cite{garfinkle}.

On the other hand already in~\cite[Thm.\ 1]{gt} it was observed that the gt-setting only
allows for a limited range of applications: The support of the Riemann tensor of a nondegenerate
gt-regular metric can only be concentrated to a submanifold of codimension of at most one. Hence
thin shells of matter can be described in the gt-setting while cosmic strings, and 
point particles cannot be covered. This fact provides a
strong motivation for a ``generalised curvature framework'' whose basis we recall now. 

Again due to the fact that the generalised framework allows to proceed componentwise 
we may define without any obstacle (see~\cite[Def.\ 6.1]{genprg}).
\begin{df}
Let $\g\in\gs^0_2(M)$ be a generalised metric then the Riemann and Ricci tensor as
well as the generalised scalar curvature are defined by the usual formulae.
\end{df}
Moreover, we have the following basic consistency with the smooth theory: If one (hence any) 
representative $\g_\eps$ of a generalised metric $\g$ converges locally uniformly together
with its derivatives up to order $2$ to a vacuum solution of Einstein's equations 
(which then necessarily is a ${\mathcal C}^2$-metric), then the Ricci tensor
of $\g$ is associated to $0$. For details see~\cite[Sec.\ 6]{genprg}.

\section{Compatibility}\label{compatibility}

So far we have described the distributional and the generalised setting in parallel.
A major question, however, is the compatibility between these frameworks, which we are going to
discuss now: Given a nondegenerate gt-regular metric $\g$, 
we have at our hands two ways to compute its curvature. The first one is to proceed within 
the gt-setting to compute $\Riem[\g]\in{\D'}^1_3(M)$, while the the second one consist of
embedding $\g$ into $\gs^0_2(M)$ and to calculate the curvature $\Riem[\iota(\g)]\in\gs^1_3(M)$
of the smoothed metric $\iota(\g)$ within the 
generalised framework. Naturally the question arises, whether these two procedures lead to 
the same result. A more precise formulation of this question is whether
the generalised curvature tensor $\Riem[\iota(\g)]$ of the embedded metric is associated to the 
distributional curvature tensor $\Riem[\g]$ of the original metric, 
i.e., whether the folowing diagram commutes.
\[
\begin{CD}
\GT\ni\g @>\iota>>[\iota(\g)]\in\G\\
@V\mbox{gt-setting}VV @VV\mbox{$\gs$-setting}V\\
\Riem[\g]@<\approx<<\Riem[\iota(\g)]
\end{CD}
\]

Since we are only interested in convergence results, it suffices to work locally:
denote by $g_{ij}$ the local components of $\g$ and write $g^\eps_{ij}$ for their
smoothings, i.e., $g^\eps_{ij}=g_{ij}*\rho_\eps$, where $\rho$ is a mollifier as in Sec.\ 
\ref{prerequisites}, and denote the resulting metric by $\g_\eps$. 
Now the question of compatibility 
may be rephrased as a question of stability: does the convergence of $\g_\eps\to\g\in
\GTconv$ for all $p<\infty$, which follows from the standard properties of convolution, 
imply the $\D'$-convergence of the curvature.

Indeed in~\cite{gt} and in~\cite{LeFM} several stability results have been provided. For our
purpose it will be sufficient to recall the following one\footnote{But see the discussion at the
end of Sec.\ 5 in~\cite{gt-paper}.}.
\begin{thr}\label{stability}(\cite[Thm.\ 4.6(2)]{LeFM}) 
If a sequence of nondegenerate gt-regular metrics converges in $\Hl^1$ and the sequence of
its inverses convergences in $\Ll^\infty$, then their distributional Riemann and Ricci tensors
converge in distributions.
\end{thr}

We now see that this result is not strong enough for our purpose.
Indeed ~(\cite[Prop.\ 4.8]{gt-paper}) the inverse $\g^{-1}_\eps$ again converges in 
$\GTconv$ for all $p<\infty$, which falls short of implying the assumptions of the Theorem. 
A positive answer to our question is, however, provided by~\cite[Thm.\ 5.1]{gt-paper} under an
additional assumption (called stability) which guarantees that the smoothing 
$\g_\eps$ of $\g$ obtained via convolution with mollifiers from as suitable class (called 
admissible) indeed provides a generalised metric in the sense of Definition~\ref{genmetric} 
(see~\cite[Sec.\ 4]{gt-paper} for details). An analogous statement is provided for
the Ricci tensor and the scalar curvature (\cite[Cor.\ 5.2]{gt-paper}) so that we 
may give the following precise statement. 
\begin{thr}
Let $\g$ be a nondegenerate, stable (see~\cite[Def.\ 4.5]{gt-paper}), gt-regular 
metric and $\g_\eps$ be a smoothing of $\g$ obtained via convolution with an
admissible (see~\cite[Lem.\ 4.3]{gt-paper}) mollifier. Then we have
\[
 \Riem[\g_\eps]\approx\Riem[\g],\ \Ric[\g_\eps]\approx\Ric[\g],\ \mbox{and}\ R[\g_\eps]\approx R[\g].
\]
\end{thr}

\section*{Acknowledgement}
The author wishes to express his warmest thanks to Stevan Pilipovi\'c and his team 
for the kind invitation to and the hospitality during the 12th Serbian Mathematical Congress.

\end{document}